# A Never-Ending Story: Revisiting Requirements Major Misunderstandings

**Julio Cesar Sampaio do Prado Leite**[0000-0002-0355-0265]

Instituto de Computação – Universidade Federal da Bahia
Rio de Janeiro – RJ – Brasil

`julioleite@ufba.br`

***Abstract.*** *A magic medallion is central in Michael Ende's novel, and it is depicted as two snakes biting each other, in a loop. Folk tale says that the design of the medallion changed for the Wolfgang Petersen's movie, depicting an even deeper image of infinity. The medallion turned out to be an icon for the story's fans. This paper will unleash a broad view of the realm of requirements and requirements engineering, comparing it to Percival's quest for the Holy Grail. Using literate and pop metaphors the paper posits that requirements engineering is an education process, which must be performed with transparency. Historical misunderstandings of requirements are reviewed, pitfalls to avoid are signaled and new trails to be built are proposed.*

***Resumo.*** *A medalha mágica é central no romance de Michael End. Esta medalha mostra duas cobras mordendo uma à outra, em um enlace. A crença popular diz que o desenho da medalha mudou para o filme de Wolfgang Petersen, ressaltando, na imagem, uma sensação de infinito ainda maior. Essa medalha tornou-se um amuleto para os fãs da estória. Esse artigo irá brotar uma visão ampla no campo de requisitos, comparando-a à busca de Percival pelo cálice sagrado. Utilizando metáforas acadêmicas e da cultura popular o artigo afirma que a engenharia de requisitos é um processo educacional, que deve ser feito com transparência. Equívocos históricos sobre requisitos são revistos, armadilhas a serem evitadas são apontadas e novos caminhos a serem construídos são propostos.*

## 1. Introduction

This paper is a reviewed edition of an earlier publication in the WER 22 – 25th Workshop on Requirements Engineering [Leite 2022]. As per the title, it treats misunderstandings about requirements. In this context, the major misunderstanding is akin to the conception that software production is marked by well-defined steps. In this conception, usually, the step for producing a requirements document is the first one.

Worse is the fact that this view is commonly taught in universities and in training courses, as of today. In a Dagstuhl workshop held in 2008, Brooks [Brooks 2009] explicitly warned about this unfortunate situation.

This misunderstanding generated and continues to generate negative impacts, like:
- A. critical problems in the final product,
- B. waste of resources in the production cycle,
- C. lack of confidence in software engineering, and
- D. lack of confidence in requirements engineering.

The title of the article is just to stress this point: the requirements artifact is a never-ending story, as well as the field of requirements engineering itself. This paper is careful in explaining what that means on a broader view of the requirements engineering discipline.

However, other major misunderstandings can be listed, like:
1. failure to understand the concept of context,
2. the failure to understand the intertwining of the several levels of design,
3. the rush to formalize, or cast in stone, when only partial semantics are available,
4. the failure to understand that by the end of the day there will be a running code to fulfill the requirements,
5. the reliance on pictures rather than models, forgetting the real meaning of analysis,
6. the failure to understand that the requirements is a result of a perennial complex political negotiation, so more than just client needs, and
7. the illusion that a complete set of requirements is just the result of good engineering.

These misunderstandings contribute to the negative impacts listed above. As such there is one major misunderstanding and seven others that contributes to the problems about requirements.

These **1 + 7** misunderstandings help contextualize the complexity of the discipline of requirements engineering. However, by equating requirements engineering to education, an even more complex scenario will be uncovered. As such, the Arthurian legend of Percival [Furtado 1992] and his quest for the Holy Grail will be a useful metaphor, for the understanding the limitations of the field.

Notwithstanding, requirements can and must be engineered. This paper posits that requirements engineering is an educational process, which must be performed with transparency. The quality of transparency aims to enhance the collaboration of different sorts of actors in each context, thus allowing for a wider participation in the process of software construction.

The text points out the already available knowledge that supports this statement, and also points out blanks that must be better studied.

The paper is organized as follows: Section 2 deals with the concept of requirements evolution, which is the major misunderstanding; Section 3 details the other 7 misunderstandings; Section 4 reviews the concept of transparency; Section 5 outlines the education metaphor. Section 6, the conclusion contextualizes the vision with other work, points out what and how there is a gain from the education metaphor with transparency, and proposes that new trails and bridges be built.

## 2. 1 – One Major Misunderstanding

First, before starting this Section, let me be clear. The viewpoint of the paper is not new, since several researchers in both software engineering and requirements engineering have pointed out the fact that requirements evolve. However, overall, this understanding is not widespread.

In The Neverending Story [Ende 1997], the medallion has the power to grant wishes. Portrayed in Wolfgang Petersen's movie [NeverEnding 1984] as two engulfing snakes, the medallion gives the idea of an infinite loop, so it portrays the idea that the story never ends. As the medallion has the power to grant wishes, who possess it may alter the state of the world, inventing a new story. However, the novel [Ende 1997] entails that the new story has only the owner of the medallion as its creator. So, at the time of creation of a new story, it will be as the creator wishes.

In Inception [Inception 2010], Christopher Nolan proposes that a new story could be created in someone dream, but in Nolan's conception, the plot of the story could be challenged by other participants of the dream that may come from one's unconscious or from other joint dreamers. Nolan's script touches issues that the field of Consciousness [JournalofConsciounessStudies 2022] has been concerned. A designed dream in Nolan's conception is an infinite space where a virtual world may be brought up. Differently from Ende [End 1997], Nolan [Inception 2010] does take in account conflicts in the proposition of a new story.

So, what those two works have to do with requirements evolution? If you believe that requirements is a story to be enacted by a machine, you have to consider that it could be rewritten as long as the writer wishes; like a new wish to the medallion. On the other hand, a new story delivered as a dream [Inception 2010] could also be written and rewritten as one wishes. However, in Inception [Inception 2010], you must be aware that, in the new story, characters may behave as they wish – if you dream with others, or if the dreamer, unconscious, act in an unforeseen way –, this is of particular interest for being aware of software intruders.

Using the story metaphor, a requirements artifact may be rewritten several times [Ende 1997], and the characters of the story may have behaviors different than the ones planned for them, and other writers could interfere in the story. In the worst case, with all the unplanned behaviors and with interference from others, the complexity of the resulting story is unbounded.

Welcome to reality. So, if the discipline of requirements engineering fails to see that the requirements artifact will change; problems will arise. The point is not when it will change, but that it will change. Lehman's definition of an E-type software states that it addresses a problem in the real world and knowledge about them cannot be absolute or complete [Lehman 1995]. The characterization of an E-type software has led to the Lehman's laws of software evolution [Lehman and Ramil 2001], which portray software as a changing artifact. Berry [Berry 2002] points out that, as a feedback system, an E-type software ends up changing its own requirements.

However, to build something, functions, and qualities [Chung and Leite 2009] need to be stated. As such the requirements artifact should be stable enough for planning the construction or thinking about its architecture. At this crucial point, software engineering, overall, still lacks well established anchors. It is incredible that the 1998 IEEE standards for requirements documents [IEEE Standards 2022][1], and the version 3.0 (2014) of the Guide to the Software Engineering Body of Knowledge [IEEE CS BofK 2022] are still tied to the phased oriented view of software construction. In the case of the Guide, it has one chapter for software construction (Chapter 3) and one for software maintenance (Chapter 5).

Several authors and educators in software engineering preach that is possible, with proper investment, to come up with a requirements artifact that is complete enough to build the right product. As such, methods have been proposed to try to write the most possible complete story before construction of the software. Many of the worst-case stories of failure in software production come from this strategy. Of course, that this is not new, different proposals for software processes came about, exactly, to answer this point [Boehm 1988] [Basili and Turner 1975] [Smith 1991] [Beck 2000] [Ebert et al. 2016]. In particular, the introduction of the concept of agile development [Beck 2000], and the practice of being even more agile, by shortening the time of deployment with DevOps [Ebert et al. 2016].

It is interesting that, in the early 80's, Davis [Davis 1982] wrote about the different strategies of requirements determination based on the available knowledge. Davis recommends that full investment in writing a "complete" story be performed only when there is enough knowledge about what is needed.

Failure to understand that requirements evolve, and evolve in various times, due to diverse types of changes, leads to negative impacts, as seen in Section 1 – Introduction. The literature has accounts of the first (A), "critical problems in the final product" [Risks Digest 2022], [Charette 2005] and the second (B), "waste of resources in the production cycle" [Breitman et al. 1999] [Berry et al. 2010] problems. For the other two problems (C) and (D), there is less literature.

The "lack of confidence in software engineering" has been addressed in two key-notes presentations in conferences; one was by James Coplien at the SBES 2001 and the other by Edward A. Lee at SOCA 2011. Be aware that there is a 10-year distance among the two keynotes, what was understood of what they said, is that software engineering failed in delivering what was expected: robust software. The point is not a discussion if they are right or not, but there is a part of world that believe that soft-ware engineering did not deliver what they thought it should. This is even worse, with the emergence of artificial intelligence-based applications, which are commonly referred as "algorithms" instead of "software".

By the same token the "lack of confidence in requirements engineering" is the feeling shared by some, that requirements engineering does not fullfil what it promised. Even being more than fifteen years old, the WER 2006[2] panel is representative. One participant at this WER edition, Suzana Oliveira, a practitioner, mentioned the term

---

[1] Reaffirmed in 2009.

[2] https://web.archive.org/web/20061106180520/http://www.ime.uerj.br:80/~vera/WER06/

"*piloto de word*" (Portuguese), which means "word driver (user)," in the sense of one's who uses word as the main tool for work. She reported that the sentence is used by implementers, coders, who disregard the job of those who write documents. Others complain about the time and volume of documents produced, which, sometimes, are not used when coding is altered, because, for instance, of the lack of proper traceability. There are also complaints that the requirements produced, although taking time and resources, did not tackle, for instance non-functional requirements. In particular, the case of non-functional requirements which are still overlooked by industrial practice [Habibullah and Horkoff 2021]. Again, promises not delivered. My conclusion is that the lack of understanding of software construction as an evolving process leads for such undesirable situations.

Back in 1997, the concept of requirements baseline [Leite et al. 1997] addressed the issue of evolution or co-evolution. The idea is centered upon a baseline as a reference, which is in constant evolution. Such an idea requires a powerful configuration manager that works addressing versions both at maintenance time with versioning (meaning the act process of the PDCA$^3$ cycle) and the development time with configuration of parts (meaning the do process of the PDCA cycle). To enact such a baseline, both versioning and configuration: intertwine through time, and use traces among versions as well as among parts (components). The crosscutting of those traces requires a sound implementation of the baseline configurator.

## 3. + 7 - The Other Seven Misunderstandings

Besides the failure with respect to evolution, other seven misunderstandings contribute to the four negative impacts seen at the Introduction. Below, each one of the seven misunderstandings is detailed.

### 3.1. Context

Early works on requirements relied on specifications, which, in general, did not consider the representation for context information. However, this failure of dealing with context has been being dealt by interdisciplinary communities, as in the Springer CONTEXT Series [Akman et al. 2001] [Bella and Bouquet 2019].

Others do not mention the notion of Universe of Discourse, which was coined in the field of database and reflects the context where the data will interact. Some work just faintly mentions the world outside of a software system, like for instance: the notion of interface brought up by use cases, while the concern is the interface of the software system and not its environment.

Jackson [Jackson 2006] specifically noticed this problem and proposed the use of what he named "Problem Domain". The name chosen by Jackson is not the best one, since it makes a confusion with the concept of Domain Knowledge, which may crosscut several environments. However, Jackson provided a clear description that there is a world, part of the real world, which contextualizes requirements to be fulfilled by a machine, composed of hardware and software.

---

[3] PDCA was made popular by Dr. W. Edwards Deming, who is considered by many to be the father of modern quality control; however, he always referred to it as the "Shewhart cycle"." (PDCA entry in Wikipedia)

## 3.2. Intertwining of Design Levels

When dealing with complex systems with distinct levels of abstractions, sometimes, the design of one system is the definition of another system that will have to be designed. For instance, when designing a braking system for a car, engineers produce a design which will be used as definition for the construction of its software part. Maher [Maher 1990] uses the idea of design formulation, design synthesis and design evaluation and mentions their recursive interaction, which is a way of dealing with distinct abstract design levels.

This intertwining of distinct levels of abstraction [Kramer 2007] does generate much confusion, mainly when the context is the staged software process. It is interesting to note that the i* language [Yu 1994] is one of the very few requirements languages that explicitly supports this intertwining at a fine grain. One goal can be refined by tasks and a task can be decomposed into goals.

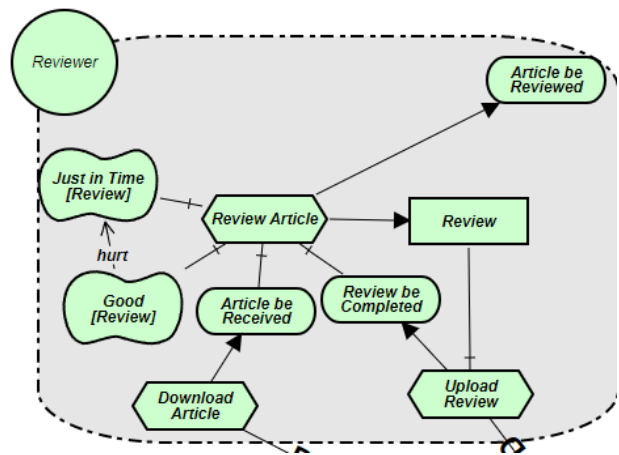

**Figura 1. i* distinct levels of abstraction**

In i*, the refinement of a goal into tasks, uses the means–end relationship, which is not a decomposition with an AND semantics, but a refinement based on the OR/XOR semantics. As in the example of Figure 1, the more abstract goal is "Article be Reviewed", which is an end, which has a means the task: "Review Article". It happens that "Review Article" is yet at a level of abstraction that needs that different goals be accomplished, such as the softgoals: "Just in Time" and "Good"; as well as the goals: "Article be reviewed" and "Review be Completed". As such, we have distinct levels of abstraction in the intertwining of goals and tasks, keeping in mind that a task is a mean to achieve a goal, so it is less abstract.

Notwithstanding, this is one of the semantics that is hard to get from the language, and several papers reporting on i* modeling do not use this as intended by the i* inventor [Yu 1994]. In i* the means-end relationship is crucial to bring variability to the requirements. This difficulty led to a variation of the language [Dalpiaz et al 2016] in its version 2.0. This move made it easier the use of i* but has the burden of loosing the capability of intertwining different abstract design levels.

### 3.3. Formalization

No one disagree of the need for formal descriptions. Fighting ambiguity [Berry and Kamsties 2004] and being able to automate a give description is central to the effort of requirements, however, let's look at this example given by Jackson and Zave [Zave and Jackson 1997].

"*Able: Two important basic types are student and course. There is also a binary relation enrolled. If types and relations are formalized as predicates, then:* $\forall s \ \forall c$ *(enrolled(s, c) =>student(s)* $\wedge$ *course(c))*.

*Baker: Do only students enroll in courses? I don't think that's true.*

*Able: But that's what I mean by student!*" [Zave and Jackson 1997].

They use this example to stress one of the dark corners of requirements engineering, which is grounding formal representations in reality. So, without a proper ontology [Breitman and Leite 2003] [Antonelli et al. 2012] to ground terms, formalization is not delivering what is promised. Several proposals based on formal descriptions leave open the grounding of terms.

Another important issue about formalization is the representation of qualities, or non-functional requirements. Chung et al. [Chung et al. 2000] proposed the use of softgoals to stand for qualities, as seen in Figure 1. The term was well chosen, given that qualities/softgoals are not satisfied but satisficed. Satisficing was a term coined by Hebert Simon [Simon 1982] is his work on behaviorism to acknowledge the fact that there are situations were optimal solutions are not attainable. The use of the term in requirements engineering is proper, since achieving a consensus over ways of achieving a quality is not clear cut among stakeholders.

### 3.4. Code is King

Over and over, we read and discuss requirements without realizing that by the end of the day, there will be running code. The phased view of software engineering has created an unnecessary gap from requirements to code. The work of Beck [Beck 2000] made agile development widely accepted [Hoda et al.], bringing requirements close to code. The "on-site customer" practice [Beck 2000] is key towards reducing the code-requirements distance. This is being brought to another dimension with the concept of DevOps [Ebert 2016]. An important contribution to bridging this gap, is the increasing role of Open Software [do Prado Leite 2018].

On the other hand, the work in requirements monitoring [Fickas and Feather 1995] [Lemos et. al. 2013] is paying attention to the problem, by providing a way to compare, at running time, if requirements are being met. In 1988, Bill Curtis et al. published a study that showed the organizational distance from customers and coders, and how this has impacted the quality of the process [Curtis et al. 1988].

### 3.5. Models

Why are models needed? The answer is because with models [Mylopoulos 1988] it is possible to look ahead in the sense that analysis be performed, which makes possible to preview how an artifact will behave without having to produce it. Models are a key element in engineering design, and they are now supported by a variety of software that help the engineer in designing the model and in analyzing them. General products, like Mathematica [Wolfram 2022] and specific ones, such as the products by ANSYS [Ansys 2022] are being used by thousands of engineers.

However, in software engineering, in general, and in requirements engineering therefore, some misunderstand the concept of models and rely on pictures, which may or may not explain what is intended. An instance of this is the reliance on so–called UML models, which in general are more concerned on syntax details of arrows and boxes instead of providing a platform for analysis and simulation like the engineering models do. Notwithstanding, model analysis is provided by both academia [Jackson 2022] and industry [IBM Rational 2022].

On top of that, there is a clear misunderstanding of the word "analysis" in general [Leite 2005]. The word is still commonly used to mean requirements elicitation, instead of the meaning engineering uses, that is ways of confirming the correctness of the model. It is proper to use the "analysis" term in its strict sense, given that the requirements process is composed of four main inner processes: elicitation, modeling, analysis, and managing. In analysis, V&V (verification and validation) strategies provide the requirements engineer with support to achieve a quality construction process.

## 3.6. The Political Game

The understanding that the RE process follows a constructionist perspective [Ramos et al. 2005], in which knowledge is created by a social process with an interaction with the environment, puts in evidence the role of stakeholders' emotions beyond the usual perspective of functional, structural, and economic RE aspects. Looking from the lens of political ecology [Bergman et al. 2002], the RE process is a political decision process, where the specification must be agreed–on through the enactment of organizational power, thus requiring political techniques to address the solution–problem space both technically as politically. Recognizing the political game, a proposal of a framework [Milne and Maiden 2012] points out a RE oriented structure to: describe power, diagnose power, and exercise power.

However, focusing on just human beings as the providers of information as to base the elicitation is a faulty procedure. Since one of the first steps in requirements engineering is trying to identify the information sources [Leite et al. 2007], from which the knowledge needed to construct the models will be elicited, usually there is a central focus on humans as the primary source. It happens that information sources are not only human beings, but there is also a plethora of information sources ranging, from laws, environment, hardware, books, regulation, and software.

It is important to mention the increasing attention to the importance of legal compliance [Engiel et al. 2017], which reflects societal concerns [Europa 2022] with software.

For instance, suppose you need requirements for updating a set of new sensors on the control of lighting. Information sources will be humans that desire to explore new capabilities of the sensors, but also the existing software, the environment where the sensors will be, the software that the vendor supplies, and so on. However, what seems to be a crucial point, not always, completely, understood, is that in the design or re-design of a system, there will be different interests at play, and it is not just a matter of understanding that there are different viewpoints [Leite 1996] but being able to negotiate these different opinions [Jureta 2009] as to address the solution–problem space both technically as politically.

### 3.7. Percival's Quest

As stressed in Section 2, above, requirements evolve and co-evolve with software production. Thus, it is a mistake to say that the requirements is complete, due to the notion of completeness fallacy [Arango and Freeman 1988], which is: the requirements is inherently incomplete in the context of E-type software [Lehman 1995].

However, several books and articles persist in claiming that a requirements document must be complete. Are they wrong? No. Despite the completeness fallacy, keep in mind: a requirements document should be socially acceptable by the stakeholders, as the source for answers to questions about the software product.

The completeness, in the requirements engineering sense, is partial but should be sufficient. This is a hard–to–understand concept, for those who like to have all the semantics settled upfront.

It is also the case that there is literature demanding that a requirements document be consistent up front. However, the requirements engineering research community has discovered that not only it is infeasible to maintain consistency at all times, but it may be desirable to allow inconsistences in the requirements process before committing to a agreed upon specification. The use of different viewpoints as mentioned above is an example of the positive aspects of allowing inconsistencies in the process. Hadar et al. [Hadar et al. 2019] examines the consistency conundrum in the realm of practitioners.

## 4. Transparency

Software is considered transparent if it makes the information it deals with transparent – information transparency – and if it, itself, is transparent, that is it informs about itself, how it works, what it does and why – process transparency–.

"*Transparency is a concept related to information disclosure, having been used in different settings, mostly related to the empowering of citizens with regard to their rights. The paper argues that, in order to implement transparency, society will need to address how software deals with this concept*." [Leite and Cappelli 2010].

In an effort improve our understanding of transparency, the software transparency research group[4] used the NFR framework [Chung et al. 2000] as basis for representing the quality of transparency. The transparency framework posits that are 5 softgoals that help transparency: "Accessibility", "Usability", "Informativeness", "Understandability", and "Auditability. Figure 2 shows the contribution of these softgoals to transparency.

---

[4] http://transparencia.inf.puc-rio.br/

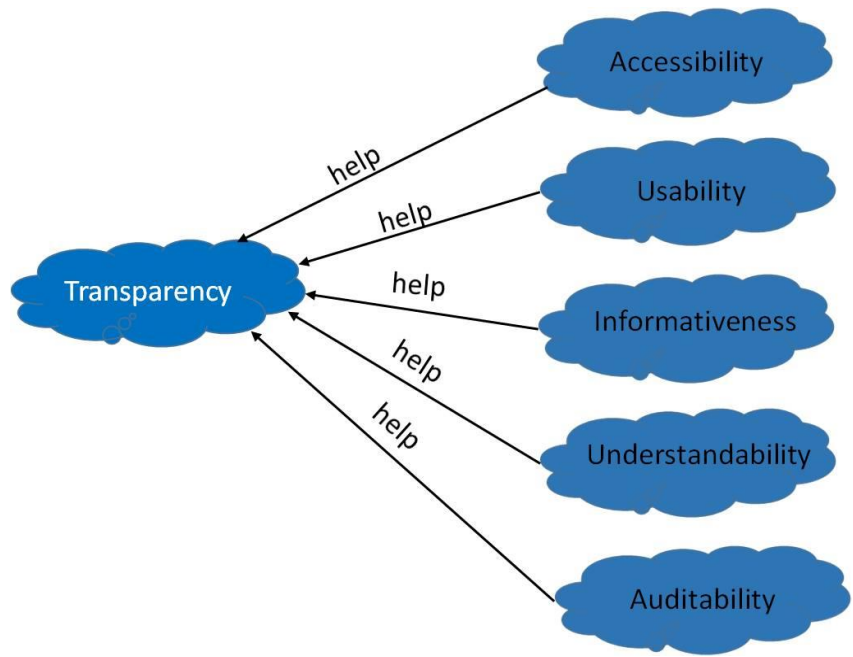

**Figure 2 Contributions to Transparency**

Each of these 5 softgoals is in turn helped by other 28 softgoals, totaling 33 softgoals helping transparency. The semantics of the help, as defined by Chung et al. [Chung et al. 2000], says that a softgoal contributes in a positive manner towards another softgoal, which does not depend on the contribution, but benefits from it. So, the contribution relation semantics should not be confused with decomposition or specialization.

How does transparency relate to requirements? To answer this question a quote from Professor John Mylopoulos is key: *Transparency is an interesting quality because it makes it necessary to attach requirements models to software*.

As such, making requirements more transparent and attaching them to software (code) contributes to (help) the overall quality of requirements, and makes explicit the options taken by the requirements team, which helps avoiding the other 7 misunderstandings. As seen in Section 2, the requirements baseline concept considers the software construction space a traceable space, see Figure 3 as in [Leite et al. 1997].

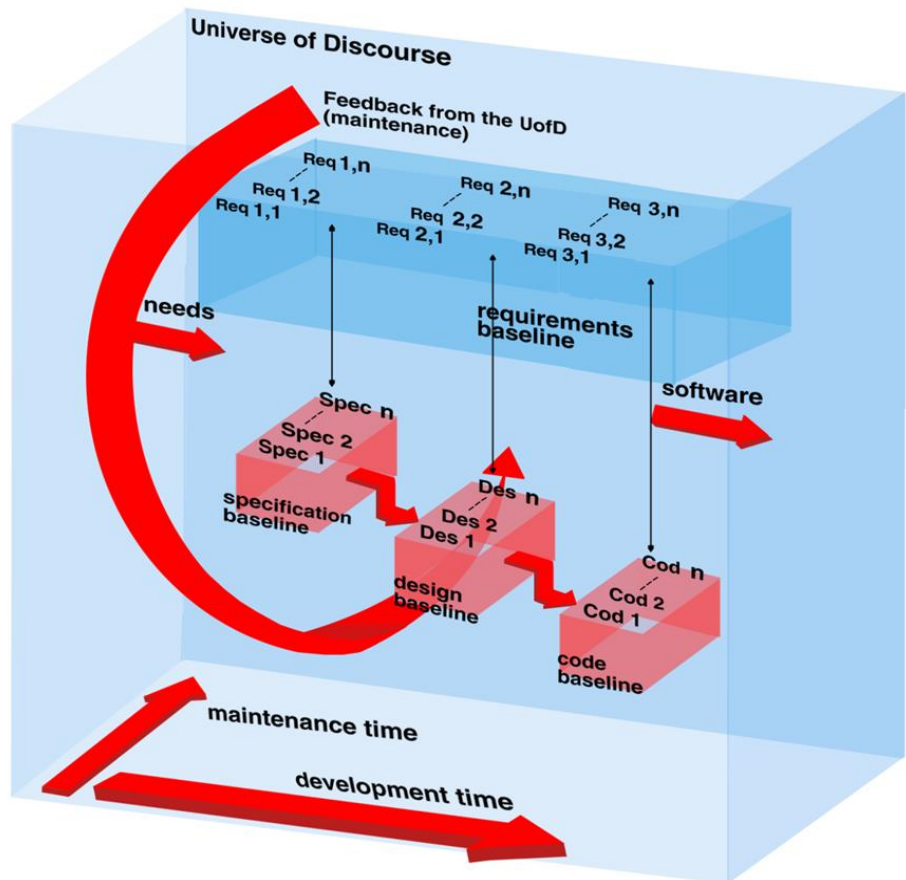

**Figure 3 The Software Construction Space [Leite et al 1997]**

Here, we should bear in mind that requirements always exist in a software. They may be implicit or explicit (transparent), so even when only the source code is available, we still have the requirements, although implicitly.

Transparency is just one of the several social demanded qualities, as well as safety, security, privacy., among others. Research in socially responsible software is an open field, Cysneiros and Leite [Cysneiros and Leite 2020] have proposed to face it from the point of view of non-functional requirements.

The next section, Section 5, sums up the factors 1 + 7 that contributes to requirements being a never-ending story. The concept of transparency and the metaphor of education are used to posit what the community has learnt, but, yet, failed to communicate to the world. Keep in mind: not only the artifact produced is prone to evolution, but the field itself is a never-ending story.

## 5. Education

[Britannica 2022] gives a general definition of education, which is used in this paper. *"Education, discipline that is concerned with methods of teaching and learning in schools or school-like environments as opposed to various nonformal and informal means of socialization (e.g., rural development projects and education through parent-* *child relationships). Education can be thought of as the transmission of the values and accumulated knowledge of a society."* [Britannica 2022]. The use of parts of this definition, in the following paragraphs, is marked by the italic font.

Education as a discipline has a long tradition. All major universities worldwide have a department or a faculty of education. Thousands of conferences worldwide discuss the theme, and a large number of books is available about the subject.

The educational process must be at hand if the goal of getting someone educated is to be achieved. If such a process could be summarized, it is proper to say that three main actors are involved: producers of learning material, educators, and students, with the goal that students be educated.

In an education system, teachers use resources already available, learning material, commonly books. Using books and other material, teachers instruct students according to the common ground knowledge contained in these supporting artifacts. On the other hand, other actors in the education process, usually called authors, handle the writing of these supporting learning artifacts. Authors are those who cast the knowledge into artifacts These authors are often educators, with profound knowledge of the discipline in question. Teachers use a sort of different strategies as to pass the knowledge from the supporting material to students, including procedures for feedback control, also known as exams.

Let's try to argue about the similarity of an education process and a requirement engineering process. In a requirement engineering process, the leading actors are the requirements engineers. Requirements engineers must elicit knowledge, so they must learn. Requirements engineers must model what they learned, such that others may use this knowledge, so they must produce education artifacts. Requirements engineers must communicate with software developers as to explain the artifacts produced, the requirements specification, as such they should educate software developers about the requirements.

The similarity has limits, since should educate is on the limits of *nonformal and informal means of socialization* among stakeholders[5] in the software construction. The point is that in the role of a requirements engineer, a stakeholder, an actor, should handle *the transmission of the values and accumulated knowledge* to the other stakeholders, even if one actor has more than one role.

Let's look at the task of eliciting the knowledge. If this metaphor is to be followed, we are talking about someone that could author a book! So, there is a heavy responsibility here, and that is why requirements elicitation is hard. Of course, that the difficulty of the learning process is proportional to the difficulty of the universe of discourse at hand, to the already available materials produced by others and to the earlier experience of the learner, an actor on the role of a requirements engineer.

How about modeling? Authoring an educational book is known to be difficult, not only knowledge is needed, but a special skill as to provide the learner with good explanations, examples, exercises. In these cases, of course, that quality control is fundamental. No one would like to learn from a book with defects. Usually book writers use natural languages, and here is where the requirements engineer has an edge. Modeling languages [Mylopoulos 1988] will allow the professional to write models that could be analyzed, but again just knowing what was elicited isn't enough, one must master the representation language of the model[6].

---

[5] In a software project, stakeholders are a set of actors that sometimes may play more than one role [Hadad 2017]. So, it is the case that in some cases the coder is the requirements engineer, or the requirements is the user, or the coder is the user, and so on.

Once the learning material is available, who will teach the students? Note that here the role of the requirements engineer is reversed, that is, now the engineer is the teacher. As such, the engineer is responsible for *the transmission of the values and accumulated knowledge* to software developers and other interested stakeholders about what the requirements is all about. In general, this part is missed in the requirements engineering process, since there is a wide acceptance that the requirement specification models produced are enough to explain what was learned. Here is where transparency comes handy. If the requirements specification is built considering the transparency quality, it will be easier to the stakeholders, even a citizen, to understand it. Note the use of the word citizen, in the sense that transparency aims to reach out all kinds of stakeholders.

How about analysis? How do educators analyze their performance and that of students? Exams is usually the feedback control for both students and educators, since evaluation, in general, is based on scores obtained by students on standard exams, but also they are a direct feedback to students on what has been learned. Requirements engineers use analysis techniques to obtain feedback, some of these techniques involve other stakeholders and are usually classified as validation techniques [Sarmiento et al. 2014], ones that feedback comes from the outside, but other type of techniques, classified as verification techniques [Sebastián et al. 2017], makes it possible that requirements engineers, by themselves, verify the written models. With respect to feedback, the availability of representation languages and proper analysis techniques is an edge to requirements engineers. However, these languages and analysis techniques are still not as popular as needed, being seldom used by the software development industry.

So, what is the purpose of using this similarity as a metaphor? What is new here? It is a belief of this paper, that there are four major advantages of studying this metaphor in more detail.

1) It makes clear that requirements engineering, as a field, and requirements documents are a never-ending story. Education is about bringing knowledge to the masses, and knowledge is being produced by research in a continuous fashion.
2) Requirements engineering is hard, if a field as old as Education is still going through revolutions, requirements engineering as a field has a lot to cover.
3) The metaphor makes it easier to see why transparency is important for requirements. More participation on the understanding of requirements, the better the requirements will be. This brings up the importance of collaboration [do Prado Leite 2018] in producing requirements.

---

[6] It is important to be aware that there is no software without requirements, even if only the code is available. As said earlier the requirements may be implicit or explicit.

> 4) The metaphor opens new ways to think about requirements engineering and to learn from the Education field, which has been exploring similar quests for a long time.

## 6. Conclusion

Let's recap the purpose of this paper: stress that evolution is key to requirements engineering, point out several factors that lead to problems in the production of requirements as a software artifact, and the proposal of a metaphor with the field of Education. The contribution relies on revising literature that deals with these identified misunderstandings and on proposing a metaphor that helps the overall understanding of the field, the profession, and of new research paths.

This paper by its own classification, is a vision paper, holds several beliefs of the author, which are justified by argumentation and, as such, is prone to bring discussions, which is just the usual goal of scientific reports.

Other work has been published discussing the area of Requirements Engineering in general [Nuseibeh and Easterbrook 2000], [Jarke et al. 2011]. The work of Jarke et al. [Jarke et al. 2011] is of special mention because it also points out some of the misunderstandings listed above. In this paper the authors point out four new principles that should give a north to Requirements Engineering, these principles are: "(1) intertwining of requirements with implementation and organizational contexts, (2) dynamic evolution of requirements, (3) emergence of architectures as a critical stabilizing force, and (4) need to recognize unprecedented levels of design complexity." [Jarke et al. 2011].

The paper has dealt with (1) in Context and Code is Key, dealt with (2) in 1, dealt with (3) in Intertwining of Design Levels and in Models and dealt with (4) in The Political Game. The paper adds to the discussion and to an overall comprehension of the field, stressing its challenges vis a vis a comparison to the field of Education. Future challenges are of varied shapes.

There is a folk story that goes like this: *A Japanese factory was struggling to find out a defect in the production of a fine mechanics product, but all the analysis performed by the engineers has failed. Due to the persistent and continuous problem the management resolved to involve all the employees in trying to solve the problem. As such, the engineers prepared a concise paper explaining the problem, when it occurred, and what was the consequence of the problem to the company. It happens that one person working as a secretary found out that the times the problem occurred where exactly the times that the fast train would pass by the factory. This was reported to the management, who passed it to the engineers. At first, they did not think it was useful, since the sensors were not detecting any discrepancies, but when they stop to look in more detail, they found out, that yes, the fast trains were causing the problem.*

This story is used over and over to exemplify why sharing information is beneficial if one wants to find problems. Open-source software development has been profiting from this philosophy of more eyeballs, and some believe it is key to their success.

As such, keeping transparency as a softgoal should provide benefits, as more and more people will have access and can understand requirements. As argued, the better exploration of the Education metaphor will lead to address the fact that requirements engineers are not taught to function as communicators. Research could help by providing ways that requirements engineers have techniques borrowed from Education [Monsalve et al. 2015]. Improving communication, a key capability to educators, will help the gap among the different stakeholders related to software. It goes without saying that the metaphor is also applied to the education of requirements engineers, as new and diverse ways of teaching the tricks of the trade are a challenge to requirements engineering pedagogy [Portugal 2016].

**Acknowledgement**

The author thanks the partial support of CNPq and CAPES. Many thanks to the RE Community at large, the IFIP 2.9 WG, PhD and Master collaborators, Co-Authors, PUC-Rio, UERJ, University of Trento, University of Toronto, University of Kaiserslautern, and University of California Irvine. Special acknowledgment to Professor Daniel M. Berry for helping to improve this revised edition.